\documentclass[aps, prb,
twocolumn,showpacs]{revtex4}
\usepackage{t1enc}
\usepackage[latin1]{inputenc}
\usepackage[english]{babel}
\usepackage{graphics}
\usepackage{graphicx}
\usepackage{dcolumn}
\usepackage{bm}
\usepackage{color}
\begin{document}
\title{A reevaluation of the coupling to a bosonic mode of the charge carriers in
(Bi,Pb)$_2$Sr$_2$CaCu$_2$O$_{8+\delta} $ at the antinodal point}
\author{J.\ Fink,$^{1,2}$  A.\ Koitzsch,$^1$
J.\ Geck,$^1$  V.\ Zabolotnyy,$^1$  M.\ Knupfer,$^1$ B.\
B\"uchner,$^1$ A.\ Chubukov,$^3$ H.\ Berger,$^4$}
\affiliation{$^1$ Leibniz-Institute for Solid State and Materials
Research Dresden, P.O.Box 270116, D-01171
Dresden, Germany\\
$^2$Ames Laboratory, Iowa State University, Ames, Iowa 50011, USA\\
$^3$Department of Physics, University of Wisconsin, Madison, Wisconsin 53706, USA\\
$^4$Institut de Physique de la Mati\`{e}re Complex, Ecole
Politechnique F\'{e}d\'{e}rale de Lausanne, CH-1015 Lausanne,
Switzerland\\}

\date{\today}

\begin{abstract}
Angle-resolved photoemission spectroscopy (ARPES) is used to study
the spectral function of the optimally doped high-T$_c$
superconductor (Bi,Pb)$_2$Sr$_2$CaCu$_2$O$_{8+\delta}$ in the
vicinity of the antinodal point in the superconducting state.
Using a parameterized self-energy function, it was possible to
describe both the coherent and the incoherent spectral weight of
the bonding and the antibonding band. The renormalization effects
can be assigned to a very strong coupling to the magnetic
resonance mode and at higher energies to a bandwidth
renormalization by a factor of two, probably caused by a coupling
to a continuum. The present reevaluation of the ARPES data allows
to come to a more reliable determination of the value of the
coupling strength of the charge carriers to the mode. The
experimental results for the dressing of the charge carriers are
compared to theoretical models.

\end{abstract}

\maketitle

\section{\label{sec:intro} Introduction}
The dressing of the charge carriers in high-T$_c$ superconductors
(HTSCs) is still one of the most exciting topics in solid state
physics. The HTSCs are a paradigm for the transition of a
correlated system from an insulating to a metallic state. The
dressing of the charge carriers in HTSCs is likely caused by the
same interaction that gives rise to the superconductivity, hence
the understanding of the quasiparticle self-energy may help to
understand the origin of the mechanism of high-T$_c$
superconductivity. The dressing can be studied by various
experimental methods but angle-resolved photoemission spectroscopy
is the only method which gives a quantitative information on the
momentum dependence of those renormalization effects. In HTSCs
there are two important regions on the Fermi surface: the nodal
region, where the diagonal of the Brillouin zone cuts the Fermi
surface and where the $d-$wave superconducting order parameter
changes sign. This region mostly contributes to the transport
properties, particularly in the underdoped region, where a
pseudogap opens up, squeezing  the Fermi surface to a region near
the nodes. The other (antinodal) region is one where the edge of
the Brillouin zone cuts the Fermi surface. In this region, the
order parameter in hole doped superconductors has its maximum.
This region is therefore mostly relevant for the studies of the
superconducting properties. There are numerous ARPES studies on
the renormalization effects near the nodal
point,\cite{Damascelli,Campuzano,Fink} but only a few studies are
concentrated at the antinodal point \cite
{Dessau,Norman1,Kaminski, Sato, Kim,Gromko}.

In the bilayer systems the study of the antinodal point is
complicated by the bilayer splitting, which could not be resolved
for 15 years. On the other hand, only in the bilayer system of the
Bi-HTSC family the entire superconducting region from underdoped
(UD) via optimally doped (OP) to overdoped (OD) can be studied. In
the superconducting state a well pronounced peak-dip-hump
structure has been detected \cite{Dessau,Norman1}. This structure
was originally explained \cite{Norman1,ABCHU} solely in terms of a
coupling to a bosonic mode, similar to the McMillan-Rowell
explanation of the tunnelling spectra in conventional
superconductors\cite{McMillan69}. Later on, it was established
that this peak-dip-hump structure is partially caused by the
bilayer splitting \cite{Kordyuk5,Borisenko3}. By varying the
photon energy h$\upsilon$ in the ARPES experiments, and exploiting
the different energy dependence of the matrix elements for the
excitations from the bonding and the antibonding bands, it became
possible to separate the two bands \cite{Kim, Kordyuk5,Borisenko3,
Bansil, Fujimori} and to extract  the full energy- and
momentum-dependent spectral weight separately in each of the
bands. This procedure allowed the authors of
Refs.~\onlinecite{Kordyuk5,Borisenko3} to find the intrinsic
peak-dip-hump structure, and to demonstrate that the strength of
this intrinsic effect is doping-dependent, and decreases in going
from UD to OD materials.

An important characteristic of the interaction between fermionic
and bosonic excitations is the energy-dependent, dimensionless
coupling $\lambda_E$. In theories where the fermionic self-energy
depends on energy, $E$, much stronger than on the momentum
$k-k_F$, this dimensionless coupling is related to the self-energy
via $\Sigma (E) = - E\lambda_E$. It is also relevant whether the
measurements are performed in the normal or in the superconducting
state. We will label the corresponding couplings as
$\lambda_{n,E}$ and $\lambda_{sc,E}$, respectively.

If the normal state is a Fermi liquid, $\lambda_{n,E=0} =
\lambda_n$ is finite, and is often called a dimensionless coupling
constant. It determines the mass renormalization of the fermionic
quasiparticles via $m^* = m( 1+ \lambda_n)$. The coupling constant
can, in principle, be extracted from  ARPES measurements of the
quasiparticle dispersion in the normal state at the lowest
energies, however this procedure requires one to know both $k_F$
and the {\it bare} mass, $m$.  In previous analysis~\cite{Kim},
the mass, $m$, was extracted from a tight-binding model with
parameters derived from a fit of the Fermi surface and from the
quasiparticle dispersion measured along the nodal
direction~\cite{Kordyuk1,Kordyuk2}. The analysis of the
experimental data in the antinodal region yielded $\lambda_{n}
\sim 1.5$ both in UD and OD materials. This result should be
contrasted with values~\cite{Kordyuk6} of $\lambda_n \leq 1$ at
the nodal point. It is consistent with expectations as for non
rotationally-invariant systems the coupling $\lambda_n$ depends on
the position on the Fermi surface.

In the superconducting state, the measured quasiparticle energy in
the antinodal region is bounded by the superconducting gap,
$\Delta$, and it becomes an issue at which energy one extracts the
coupling $\lambda_{sc,E}$ from the data. In previous analysis, the
coupling was extracted from the self-energy measured at $|E|\geq
\Delta$. This coupling $\lambda_{sc,\Delta}$ turns out to be
larger than $\lambda_{n}$, and it also rapidly increases from OD
to UD samples ($\lambda_{sc,\Delta} \sim 8$ for dopant
concentration $0.12$).

In this communication we extend our previous analysis of the
antinodal self-energy in the superconducting state~\cite{Kim} and
show how one can extract the coupling at zero frequency
$\lambda_{sc,E=0} \equiv \lambda_{sc}$ from the ARPES data. We
find that $\lambda_{sc}$ is smaller than $\lambda_{sc,\Delta}$ and
within a certain model is also smaller than the normal state
coupling $\lambda_{n}$, in agreement with earlier calculations
\cite{Chubukov}. We show that the large value of
$\lambda_{sc,\Delta}$ and its strong doping dependence are at
least partially due to the fact that the fermionic self-energy in
a superconductor actually diverges at $|E| = \Delta + \Omega_0$,
where $\Omega_0$ is the energy of the bosonic mode. If the bosonic
mode is the spin resonance peak, its energy decreases with
decreasing doping. Then $|E| = \Delta$ and $|E| = \Delta +
\Omega_0$ come closer to each other in the UD regime, and
$\lambda_{sc,\Delta}$ strongly increases. This is consistent with
the analysis in Ref.~\onlinecite{Kim}.

Our present analysis is based on the measurements of the
quasiparticle spectral function  in the antinodal region of the
high-T$_c$ superconductor (Bi,Pb)$_2$Sr$_2$CaCu$_2$O$_{8+\delta} $
(BiPb2212) in the superconducting state. We go beyond a previous
ARPES study which has analyzed the energy dependence of the
spectral weight just at the $(\pi,0)$ point \cite{Norman}, and
study the whole antinodal region. We interpret our results in the
superconducting state in terms of model self-energy function which
is composed of two terms. The first and dominant term is due to a
strong coupling of the charge carriers to a single bosonic mode.
The second term describes a band renormalization at higher
energies and is assumed to have a Fermi-liquid form. We extract
both couplings  from the fits to the data. We used two models for
electron-boson coupling. The first model is a one-mode model for
an interaction with an Einstein boson, which is assumed to be
independent on fermions. Second is a collective mode model, in
which the bosonic spectrum in the normal state is rather flat and
incoherent, but splits into a mode and into gapped continuum in
the superconducting state due to the feedback effect from the
pairing. This second model is appropriate if the boson is a spin
collective mode of fermions. We obtain a rather good agreement
between the parameters derived from the analysis of the
experimental data using the model self-energy function and the
calculated values using the collective mode model. This yields a
strong indication that the dominant part of the renormalization of
the fermionic dispersion is due to a coupling of collective spin
excitations.

The paper is organized as follows. In Sec. II we review the two
fermion-boson models in the normal and the superconducting state.
The experimental setup is discussed in Sec. III. In Sec. IV we
present the experimental results together with the data analysis.
In Sec. V we discuss the results and compare them with other
renormalization effects studied by ARPES in solid state physics.
The conclusions of our study are presented in Sec. VI.

\section{\label{sec:model} The fermion-boson models}

The coupling of the charge carriers to bosonic excitations is the
minimum model to understand the spectral function of the HTSCs at
the antinodal point. We start with an assumption that the Fermi
energy $E_F$ is much larger than the mode energy $\Omega_0$. The
validity of this assumption for very underdoped cuprates has been
questioned recently \cite{KShen, Roesch} because there the
bandwidth is strongly reduced due to correlation effects
associated with Mott physics. Here we restrict our analysis to
near-optimally doped cuprates for which there is little doubt that
$E_F\gg\Omega_0$ since $E_F$ $\sim$ 1 eV in this case.

Both fermion-boson models have been discussed earlier in the
literature \cite {Norman, Schrieffer, Scalapino, Eschrig, ABCHU,
Devereaux}. We review them here again in order to specify the
parameters which can be derived from ARPES. We also present
several new results for the collective excitations model.

The dynamics of an electron in an interacting system can be described by a
Green's function \cite{Mahan}
\begin{equation}
G(E ,k)=\frac{1}{E -\epsilon _k-\Sigma(E ,k)}.
\end{equation}
where $\Sigma(E,k)$=$\Sigma'(E,k)$+i$\Sigma''(E,k)$ is the complex
self-energy function which contains the information on the
fermion-boson interaction, and $\epsilon _k$ is the bare
quasiparticle dispersion. Near the Fermi surface $\epsilon_k = v_F
(k-k_F)$, where $v_F = k_F/m$, and $m$ is the bare mass. It is
customary to use the tight-binding form for $\epsilon_k$.

ARPES experiments measure the product of the spectral function
$A(E ,k)$, the Fermi function, and a transition matrix element,
convoluted with the experimental resolution. The spectral function
is related to the Green's function as \cite{Hedin,Almbladh}
\begin{eqnarray}
A(E ,k)&=&-\frac{1}{\pi}ImG(E ,k)\nonumber\\
&=&-\frac{1}{\pi}\frac{\Sigma''(E ,k)}
{[E -\epsilon _k-\Sigma'(E ,k)]^2+[\Sigma''(E ,k)]^2}
\label{26_1}
\end{eqnarray}
For  $\Sigma=0$, i. e., for the non-interacting case, the spectral
function $A(E,k) = \delta (E - \epsilon_k)$.

For the description of the spectral function in the
superconducting case, two excitations have to be taken into
account: the electron-hole and the pair excitations. This
transforms the  Green's function into a (2x2) matrix,~\cite
{Nambu} or, equivalently, to the emergence of normal and anomalous
components of the Green's function. Accordingly, the self-energy
also has a normal part $\Sigma (E,k)$ and anomalous part $\Phi
(E,k)$. The two self-energies are related to $E$, the
renormalization function Z(E,k), and to the superconducting gap
$\Delta (E,k)$ via
\begin{equation}
E - \Sigma (E,k) = EZ(E,k),~~ \Phi (E,k) = Z(E,k)\Delta (E,k)
\end{equation}
In general, the superconducting gap $\Delta (E,k)$ is also a
complex function and depends on both parameters. The energy
dependence is not crucial, though~\cite{Abanov}, and we just
neglect it for simplicity, i.e., replace a complex $\Delta (E,k)$
by a real $\Delta (k)$. The momentum dependence of $\Delta (k)$ is
that of a $d_{x^2 -y^2}$ gap. In the antinodal region, the gap is
near its maximum, its momentum dependence is  weak and we will
neglect it as well, i.e., further approximate $\Delta (k)$ by
$\Delta$ and $Z(E,k)$ by $Z(E)$. The spectral function is then
given by \cite{Scalapino}
\begin{equation}
A(E,k)=-\frac{1}{\pi}Im\frac{Z(E)E+\epsilon_k}{Z(E)^2(E^2-\Delta^2)-\epsilon_k^2}.
\label{26_4}
\end{equation}
Using our definition of the coupling constant,
\begin{equation}
Z(E) = 1 + \lambda_{sc,E}.
\end{equation}

Below we consider two models for electron-boson interaction. In
the one-mode model we define the self-energy due to the coupling
to a single bosonic mode as $\Sigma' = - \lambda^b_E  E$ with
$\lambda^b_0 = \lambda^b$. In the collective mode model, we treat
the renormalization due to a distribution of bosonic modes. The
corresponding coupling constant is called $\lambda^c$.

\subsection{One-mode model}

\begin{figure}
\includegraphics[width=7 cm]{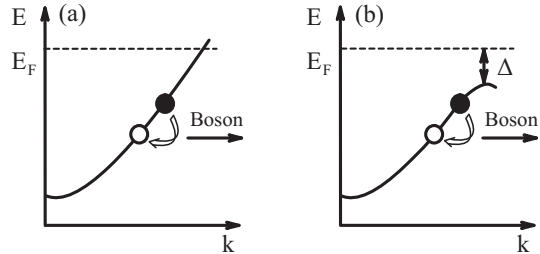}
\caption{Bosonic scattering mechanism which contributes to the imaginary part of the self-energy. (a) normal state; (b)
superconducting state.}
\label{boson}
\end{figure}

In the one-mode model, it is assumed that electrons interact with
an Einstein boson whose energy is $\Omega_0$ independent on
whether the system is in the normal or in the superconducting
state. For the normal state the mechanism leading to a finite
lifetime of a photohole is illustrated in Fig.~\ref{boson}(a). The
hole is filled by a transition from a state at lower binding
energy via an emission of a bosonic mode. Such bosonic excitations
may be electron-hole excitations, phonons, spin excitations,
plasmons, excitons etc. Relevant excitations for HTSCs are listed
in Table I together with their characteristic energies.

\begin{table}
\centering
\caption{Bosonic excitations which couple to the charge carriers together with
their characteristic energies in HTSCs}
\label{tab:1}       
%
%
\begin{tabular}{llr}
\hline\noalign{\smallskip}
system &\hspace{.1cm} excitations &\hspace{.1cm}characteristic energy(meV)  \\
\noalign{\smallskip}\hline\noalign{\smallskip}
ion lattice &\hspace{.1cm} phonons & 90\hspace{2cm} \\
spin lattice/liquid &\hspace{.1cm} magnons & 180\hspace{2cm} \\
e-liquid &\hspace{.1cm} plasmons & 1000\hspace{2cm} \\
\noalign{\smallskip}\hline
\end{tabular}
\end{table}

\begin{figure}
\includegraphics[width=7 cm]{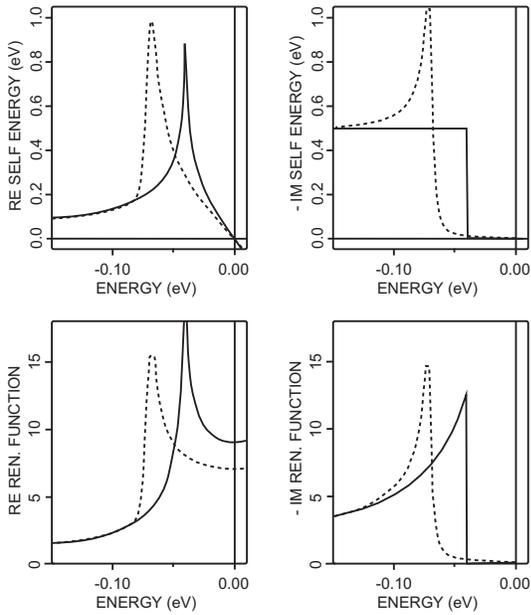}
\caption{Real (a) and imaginary (b) part of the self-energy
function and real (c) and imaginary (d) part of the
renormalization function for a coupling to a mode at $\Omega_0$ =
40 meV and a coupling constant $\lambda _n$=8. Solid line: normal
state, dashed line: superconducting state with a gap $\Delta $ =
30 meV } \label{Sigma}
\end{figure}

For a constant density of states and for the temperature T = 0,
the  fermionic self-energy  is given by\cite {Schrieffer}
\begin{equation}
\Sigma (E) = \frac{i \lambda^b_n}{2\pi\Omega_0} \int d E^\prime \chi (E^\prime)
\int d \epsilon_k G (E + E^\prime, k)
\label{25_1}
\end{equation}
where $\chi (E^\prime)$ is the bosonic propagator
\begin{equation}
\chi (E^\prime) = \frac{\Omega_0}{\Omega^2_0 - E^{\prime 2} - i\delta}
\label{25_2}
\end{equation}
In the normal state, $\int d \epsilon_k G (E + E^\prime, k) = - i
\pi \text{sgn} (E + E^\prime)$.  Substituting this into
(\ref{25_1}) and separating real and imaginary parts of the
integral, we find, for $E <0$
\begin{eqnarray}
&&\Sigma' (E) = - \frac{1}{2} \lambda^b_{n}
\Omega _0~\ln {|\frac{E+\Omega_0}{E-\Omega _0}|} \nonumber \\
&& \Sigma^{''} (E)  = \frac{\pi}{2} \lambda^b_n \Omega_0 \theta
(|E| - \Omega_0). \label{25_3}
\end{eqnarray}
$\Sigma^{\prime \prime} (E)$  is zero up to the absolute value of
the mode energy $\Omega_0$ . This is also clear from
Fig.~\ref{boson} since the photohole can only be filled when its
binding energy is larger than $\Omega_0$. At $|E|>\Omega_0$,
$\Sigma^{\prime \prime} (E)$  is a constant (see
Fig.~\ref{Sigma}~(b)). $\Sigma^\prime (E)$  shows a logarithmic
singularity at the mode energy, $\Omega_0$ (see
Fig.~\ref{Sigma}~(a)). At low energies there is a linear energy
dependence of $\Sigma'$ and the negative slope $-d\Sigma'/dE =
\lambda^b_n$ determines the coupling constant at zero energy.

In Fig.~\ref{Sigma}~(c) and (d) we have plotted the
renormalization function Z(E) for the same parameters. The real
part shows again a singularity at $\Omega_0$ and a constant value
at zero energy. This value minus 1 again determines the coupling
constant $\lambda^b_n$. Within this model the same coupling
constant can also be obtained from the measurements of the
quasiparticle linewidth at large negative energies as $\lambda^b_n
=- 2 \Sigma^{''} (-\infty) /(\pi \Omega _0)$. $\Sigma^{''}
(-\infty)$ is the step height of $\Sigma^{''}(E)$ at $E$ = $\Omega
_0$ in the one-mode model. Hence both $\lambda^b_n$ and
$\Sigma^{''} (-\infty)$ are the measures of the coupling strength
to the bosonic mode. Consequently, together with the mode energy
$\Omega_0$, both can be used to determine the self-energy in the
one-mode model.

\begin{figure}
\includegraphics[width=8.5 cm]{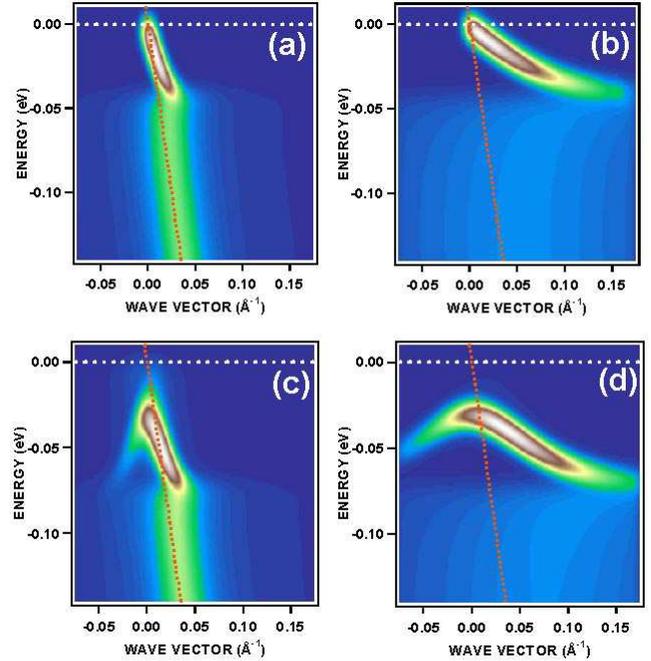}
\caption{Calculated spectral function A(E,k) for a coupling of the
charge carrier to a mode with an energy $\Omega _0$=40 meV (a) and
(b) normal state, (c) and (d) superconducting state with a
superconducting gap, $\Delta $=30 meV. (a) and (c) coupling
constant $\lambda _n$=1, (b) and (d) coupling constant $\lambda
_n$=8 } \label{AEk}
\end{figure}

In Fig.~\ref{AEk} (a) and (b) we have displayed the calculated
spectral function in the one-mode model for $\lambda^b _n=1$ and
$\lambda^b _n=8$, respectively. Compared to the bare particle
dispersion, $\epsilon _k$, given by the red dashed line, for |E| <
$\Omega_0$ there is a mass renormalization, i.e., a reduced
dispersion and no broadening, except the energy and momentum
resolution broadening, which was taken to be 5 meV and 0.005
\AA$^{-1}$, respectively . For |E| > $\Omega _0$, there is a
dispersion back to the bare particle energy. Moreover, there is a
broadening due to a finite $\Sigma''$, increasing with increasing
$\lambda^b_n$. For large $\lambda^b_n$, the width for constant E
scans is, at least up to some energy, larger than the binding
energy of the charge carriers and therefore they can be called
incoherent in contrast to energies |E| < $\Omega_0$ or very high
binding energies, where the width is smaller than the binding
energy and therefore the states are coherent\cite{Schrieffer}. The
change in the dispersion is often termed a "kink" but looking
closer at the spectral function, in particular for high
$\lambda^b_n$, there is a branching into two dispersion arms
touching each other at the branching energy E$_B$=$\Omega _0$.

The following information can be obtained from a one-mode model
spectral function $A(E,k)$ in the normal state. When performing
constant-E scans, very often called momentum distribution curves
(MDCs), one obtains Lorentzians. The maximum observed in the MDCs
determines the renormalized dispersion. By comparing this
dispersion close to $E_F$ to the bare particle dispersion one can
extract the coupling constant $\lambda^b_n$. The width of the
Lorentzians for $|E|<\Omega _0$ in this model should be determined
by the energy and momentum resolution. For $|E|>\Omega _0$ the
resolution effects are small compared to the intrinsic width, W,
and one can derive $\Sigma'' (-\infty)=-v_FW/2$ where $v_F$ is the
bare-particle Fermi velocity. From the onset of a finite intrinsic
width, one obtains the mode energy $\Omega_0$. This parameter also
can be obtained by constant-E cuts, very often called energy
distribution curves (EDCs). Looking at Eq. (2), one realizes that
for large $\epsilon_k$, i.e., far away from the Fermi wave vector,
$k_F$, the spectral function around $E\sim \Omega_0$ is determined
by $\Sigma''$. The edge of such a cut determines again $\Omega_0$.

In the superconducting state, the self-energy for $E<0$ is still
given by (\ref{25_1}), but the fermionic Green's function now has
the form
\begin{equation}
G_{sc} (E, k) = \frac{E + \epsilon_k}{E^2 -\Delta^2 - \epsilon^2_k + i \delta}
\label{25_4}
\end{equation}
Substituting this into (\ref{25_1}), evaluating the integral over $\epsilon_k$
 and separating real and imaginary parts, we obtain  for $E<0$:
\begin{eqnarray}
&&\Sigma^{'}  (E) = - \frac{\lambda^b_n}{2} \Omega_0 {\text
Re}\int \frac{d E^{'}}{\Omega^2_0 - E^{'2} -i\delta}~\frac{E +
E^{'}}{\sqrt{(E + E^{'})^2 - \Delta^2}}
\nonumber \\
&&\Sigma^{''} (E) = \frac{\pi}{2} \lambda^b_n \Omega_0 {\text Re}~
\frac{E + \Omega_0}{\sqrt{(E + \Omega_0 + \Delta) (E + \Omega_0 -
\Delta)}} \label{a_1}
\end{eqnarray}
In  Fig.~\ref{Sigma} we plot $\Sigma(E)$ and $Z(E)$ for the
superconducting state with $\Delta$ = 30 meV. A small $\delta$ has
been used to reduce the singularities. One realizes that due to
the opening of the gap the singularities of $\Sigma'$ and $Z'$ are
shifted to higher binding energies and that the edge in $\Sigma''$
transforms into an overshooting edge. At vanishing $E$, $\Sigma '
(E)$ is still linear in $E$, but due to the shift of the
singularity to higher binding energies the slope is now reduced
and given by
\begin{equation}
\lambda^b_{sc} =- \frac{\Sigma^{'} (E\rightarrow 0)}{E} = \lambda^b_n \int_0^\infty \frac{dx}{(x^2 +1)^{3/2}}~\frac{1}{1 + x^2 (\Delta/\Omega_0)^2}
\label{a_2}
\end{equation}
The reduction of the coupling constant in the superconducting
state is concomitant with a reduction of $Z(0)$ since $Z(0) = 1 +
\lambda$.

At $k=k_F$ the photohole can only be excited when its binding
energy is exactly $\Delta$, or when $Im Z (E, k_F) = - Im \Sigma
(E,k_F)/E$ is non-zero. From Fig.~\ref{boson} (b) it is clear that
for zero temperature and in the clean limit $\Sigma''$ or the
scattering rate is different from zero only when $|E| > \Delta+
\Omega_0$. This result is obtained  from an evaluation of Eq.
(\ref{a_1}). This implies that $E =-\Delta$ is separated from the
region where $\Sigma^{''} (E, k_F)$ is non-zero and therefore the
spectral function contains a $\delta-$functional peak at $E = -
\Delta$, and then it becomes nonzero at $E < -(\Delta +
\Omega_0)$.

 In Figs.~\ref{AEk} (c) and (d) we show for the one-mode model the calculated
spectral function in the superconducting state using the same
energy and momentum resolutions and the same mode energy as
before. The gap  was set to $\Delta$ = 30 meV. One clearly
realizes the BCS-Bogoliubov-like back-dispersion at the gap energy
$\Delta$ and besides this, a total shift of the dispersive arms by
the gap energy. Thus the branching energy E$_B$ occurs at
$-(\Omega_0+\Delta )$.

The renormalized dispersion is obtained from the  position of the
MDC peak of the spectral function. In the normal state, the peak
position is where the real part of $G^{-1} (E,k)$ vanishes. In the
superconducting state, there is an extra complication due to the
fact that $\epsilon_k$ is present both in the denominator and in
the numerator of the spectral function. Like in an earlier
study~\cite{Chubukov} we avoid this complication and extract the
renormalized dispersion from
\begin{eqnarray}
\epsilon_k &=& - {\text Re} Z(E) \sqrt{E^2-\Delta^2} \nonumber \\
&&= - \left(1 - \frac{{\text Re} \Sigma (E)}{E}\right) \sqrt{E^2-\Delta^2}.
\label{a_3}
\end{eqnarray}
In  conventional superconductors, the mode energy is much larger
than the gap. In this situation, Eq. (\ref{a_2}) yields
$\lambda^b_{sc} = \lambda^b_n \left(1 + O\left((\Delta/\Omega_0)^2
\log{\Delta/\Omega_0}\right) \right) \approx \lambda^b_n$.
Furthermore, the same small parameter $\Delta/\Omega_0$ also
allows one to neglect the energy dependence of $\lambda_{sc,E}$ at
$|E| \geq \Delta$, such that  $\lambda^b_{sc, E} \approx
\lambda^b_{sc} \approx \lambda^b_n$. In this situation, $Re Z (E)
\approx 1 + \lambda^b_n$, and hence the maximum of the spectral
function is located at
\begin{equation}
E=- \sqrt{\Delta ^2+\epsilon _k^2/(1+\lambda^b_n)^2}.
\label{a_4}
\end{equation}
For HTSCs, the gap is comparable to the mode energy and therefore
Eq.(\ref{a_4}) is no longer valid, and the full Eq. (\ref{a_3})
should be used to fit the dispersion. There are two key
differences with Eq. (\ref{a_4}). First, the zero-energy values
$\lambda^b_{sc}$ and $\lambda^b_n$ are different. For $\Delta =
30$ meV and $\Omega_0 = 40$ meV, i.e., $\Delta/\Omega_0 = 3/4$, we
obtain from (\ref{a_2}) $\lambda^b_{sc} = 0.74 \lambda^b_n$.
Second, the energy dependence of $\lambda^b_{sc}$ becomes
relevant. Indeed, by analyzing  (\ref{a_1}) one finds that
$\Sigma^{'} (E)$ is discontinuous at $E = - (\Delta + \Omega_0)$
and diverges as a square-root  at approaching $E = -(\Delta +
\Omega_0)$.~\cite{Chubukov} When $\Delta$ and $\Omega_0$ are
comparable, this divergence affects the self-energy already at $E
\approx - \Delta$. For the parameters that we choose, the effect
is not large:  evaluating the real part of the self-energy at $E =
-\Delta$ from (\ref{a_1})  we find $\lambda^b_{sc,\Delta} \approx
1.1 \lambda^b_{sc}$. However, the effect increases once $\Omega_0$
gets smaller.

Measuring an EDC at $k_F$ with high resolution, one would expect a
peak at $\Delta$, followed by a region of near-zero spectral
weight and a threshold of the incoherent spectral weight, which
appears at $\Omega_0+ \Delta$.  Such an energy distribution is
well known from tunnelling spectroscopy in conventional phonon
superconductors, except that there $\Omega_0$ is often much larger
than $\Delta$. At deviations from $k_F$, the peak disperses to
larger frequencies while the onset of the incoherent spectral
weight remains at $\Omega_0+ \Delta$. Once the peak disperses
close to $\Omega_0+ \Delta$, only the threshold at this energy
remains visible.

\subsection{Collective mode model}

For definiteness, we consider the model with the interaction
between fermions and their spin collective excitations with
momenta near $Q = (\pi,\pi)$. The momentum $Q$ connects Fermi
surface points within antinodal regions, and hence antinodal
fermions are mostly involved in the scattering of nearly
antiferromagnetic spin fluctuations.

The physics of electron-boson interaction is somewhat different in
the one-mode and collective mode scenarios. Like we said,  in the
one-mode formalism, one assumes that bosons are propagating
excitations with a frequency $\Omega_0$, independent on whether
fermions are in the normal or in the superconducting state. In the
collective mode model, bosons are Landau-damped in the normal
state, and their spectral function is described by a continuum
rather than by a mode. In the superconducting state, the
low-energy fermionic states in the antinodal regions are gapped,
and the continuum of bosonic states with momenta near $Q$ appears
only above the gap of $2\Delta$. In addition, the residual
attraction between fermions in a $d_{x^2-y^2}$ superconductor
leads to the development of the resonance peak at a frequency
$\Omega_0$ below $2\Delta$. In the OD regime, $\Omega_0$ is only
slightly below $2\Delta$, and  the resonance is weak. In the UD
regime, the resonance frequency decreases. In bilayer systems,
such as Bi2212, there are two resonances, in the even and in odd
channel. The resonance frequency in the even channel should vanish
at the point where the magnetic correlation length diverges. The
resonance in the odd channel remains finite at this point, and,
very likely, transforms into the gapped spin-wave mode in the
antiferromagnetically ordered state.

The self-energy within the collective mode model has been analyzed
in Refr.~\onlinecite{Chubukov} and in earlier publications. Below
we briefly review the existing results and also present several
new formulas. For definiteness, we consider the case of a flat
static susceptibility near $Q$, i.e., assume that in the normal
state, the dynamical spin susceptibility (the bosonic propagator)
can be expressed as
\begin{equation}
\chi (E, q) = \chi (E) = \frac{\chi_Q}{1 - i E/\omega_{sf}}
\label{a_5}
\end{equation}
where $\omega_{sf}$ is the typical relaxational frequency of spin
fluctuations. An advantage of using the flat static spin
susceptibility is that all computations can be done analytically.
Similar results are also obtained using Ornstein-Zernike form of
the static susceptibility~\cite{Abanov} and in FLEX computations
for the Hubbard model~\cite{manske}.

In the normal state, the fermionic self-energy due to interaction
with the gapless continuum of spin excitations is~\cite{Chubukov}
\begin{eqnarray}
&&\Sigma^{'} (E) = - \lambda^c_n \omega_{sf} \arctan{\frac{E}{\omega_{sf}}} \nonumber \\
&&\Sigma^{''} (E) = - \frac{1}{2}~\lambda^c_n \omega_{sf} \ln(1 +
\frac{E^2}{\omega^2_{sf}}) \label{a_6}
\end{eqnarray}
where $\lambda^c_n$ is the dimensionless coupling constant in the
normal state for the collective mode model. In distinction to  the
one-mode model, the self-energy in (\ref{a_6}) has no threshold,
and its energy dependence interpolates between different limits.
In particular, $\Sigma^{''} (E)$ is quadratic in $E$ at the lowest
energies (a Fermi-liquid form), and is almost flat at large $E$.
At intermediate energies, $\Sigma^{''} (E)$  is roughly linear in
$E$. The real part of the self-energy is linear in $E$ at the
lowest frequencies ($- d \Sigma^{'} (E)/d E_{E \rightarrow 0} =
\lambda^c_n$), and is flat at high frequencies. If the
relaxational spectrum of spin fluctuations is cut at some upper
cutoff,  the real part of the self-energy will start decreasing
above the cutoff.

In the superconducting state, the self-energy changes by two
reasons. First, fermionic excitations acquire a gap. Second, the
spectrum of collective excitations by itself changes as a feedback
from the gap opening. The expression for the self-energy
incorporates both effects and is given by
\begin{equation}
 \Sigma  (E)
= - \frac{1}{2} \lambda^c_n ~\int \frac{d E^{'}}{1 - \Pi
(E^{'})/\omega_{sf}}~ ~\frac{E + E^{'}}{\sqrt{(E + E^{'})^2 -
\Delta^2}} \label{a_7}
\end{equation}
where  $\Pi (E)$ is the polarizability bubble in the
superconducting state (a sum of the two bubbles made of normal and
anomalous fermionic Green's functions). This polarization operator
can be computed explicitly. We obtained
\begin{equation}
\Pi^{'} (E) = \left \{
\begin{array}{l}
\frac{E^2}{2 \Delta} D \left(\frac{E^2}{4\Delta^2}\right)~ \mbox {for}~ |E| <2\Delta \\ \\
\frac{4 \Delta^2}{|E|} D \left(\frac{4\Delta^2}{E^2}\right)~ \mbox {for}~
|E| >2\Delta
\end{array} \right .
\end{equation}
\begin{equation}
\Pi^{''} (E) =  \left \{
\begin{array}{l}
 0~ \mbox {for}~ |E| <2\Delta \\ \\
|E|~ K_2 \left(1- \frac{4\Delta^2}{E^2}\right)~ \mbox {for}~
|E| >2\Delta
\end{array} \right .
\label{a_8}
\end{equation}
where $D(x^2) = (K_1(x^2)-K_2(x^2))/x^2$, and $K_1(x^2)$ and
$K_2(x^2)$ are the elliptic integrals of first and second kind,
respectively. The expression for $\Pi^{''}$ was earlier obtained
in \cite{morr}.

We see that $\Pi^{''}$ is finite only at $|E| >2\Delta$. At $|E| <
2\Delta$, $\Pi^{'} (E)$ is positive and interpolates between zero
at $E=0$ and infinity at $|E| = 2\Delta$ (at the lowest energies,
$\Pi (E) \approx  (\pi/8) E^2/\Delta$). At some frequency
$\Omega_0$, $\Pi^{'} (\Omega_0) = \omega_{sf}$, and the dynamical
spin susceptibility $\chi_s (E) \propto 1/(1 - \Pi
(E)/\omega_{sf})$  has a pole. As a result, the gapless continuum
of the normal state splits into two separate entities: the gapped
continuum at energies above $2\Delta$, where $\Pi^{''}$ is
non-zero, and the pole (the resonance peak) at an energy
$\Omega_0$ below $2\Delta$. We see therefore that in the
superconducting state, one-mode and continuum models are quite
similar -- both describe the interaction between fermions and a
bosonic mode. The difference between the two models is in the
details, and also in the fact that in a collective mode model, the
bosonic spectrum still contains a continuum above $2\Delta$.

The location of the pole can be straightforwardly obtained from
(\ref{a_8}). For $\Delta$ = 30 meV, the mode is at $\Omega_0 $= 40
meV, if $\omega_{sf} \sim$ 26 meV. This last value is quite
consistent with earlier estimates~\cite{Abanov} Near the pole, the
spin susceptibility is
\begin{equation}
\chi (E) \approx Z_0 \frac{\Omega_0}{\Omega^2_0 -E^2 - i \delta}
\label{a_9}
\end{equation}
where $Z_0 \sim 0.77$. Apart from the residue $Z_0$, Eq.
(\ref{a_9}) describes the same propagator as in the one-mode model
(see Eq. (\ref{25_2})).

Substituting the results for $\Pi$ into the expression for the
fermionic self-energy, Eq. (\ref{a_7}), we find $\Sigma (E)$ as a
sum of two contributions. One comes from the pole and the other
comes from the gapped continuum. The generic behavior of the
self-energy is similar to what we have found for the  one-mode
model. Namely, $\Sigma^\prime (E)$ is linear in $E$ at the lowest
energies, and diverges as a square-root at approaching $-(\Delta +
\Omega_0)$ from below. Above this threshold, $\Sigma^{'} (E)$
drops to a finite value, and decreases at even larger $|E|$. The
imaginary part of the self-energy is zero below the threshold at
$-(\Delta + \Omega_0)$, diverges as a square-root at approaching
the threshold from larger $|E|$, and eventually recovers the
normal state value at highest energies. At the smallest $E$, we
found that the dominant contribution to $\Sigma^\prime (E) = -
\lambda^c_{sc} E$ comes from the mode, continuum only accounts for
about 20\% percent correction. Evaluating the integrals, we found
that $\lambda^c_{sc} \approx 0.7 \lambda^c_n$. This is similar to
what we have found in the one-mode model. At $E = -\Delta$, we
found, using the full form of the polarization bubble,
$\lambda^c_{sc, \Delta} \approx 0.75 \lambda^c_n$, which is again
similar to what we have found in the one-mode model.

For the imaginary part of the self-energy $\Sigma^{''} (E)$ and
$E$ < 0 we found
\begin{equation}
\Sigma^{''} (E) = \Sigma^{''}_A (E) + \Sigma^{''}_B (E)
\end{equation}
where
\begin{eqnarray}
&&\Sigma^{''}_A (E) = \frac{\pi Z_0}{2} \lambda^c_n \Omega_0 \frac{E + \Omega_0}{(E + \Omega_0 + \Delta)(E + \Omega_0 - \Delta)} \nonumber \\
&& \Sigma^{''}_B (E) =- \lambda^c_n \int_{2\Delta}^{|E|} dx \text{Re} \frac{E+x}{\sqrt{(E+x)^2 -\Delta^2}}~\times \nonumber \\
&&\frac{\frac{x}{\omega_{sf}} K_2 (1- \frac{4\Delta^2}{x^2})}{\left(1- \frac{4\Delta^2}{x \omega_{sf}}D(\frac{4\Delta^2}{x^2})\right)^2 + \left( \frac{x}{\omega_{sf}} K_2 (1- \frac{4\Delta^2}{x^2})\right)^2}
\label{a_10}
\end{eqnarray}
The first contribution is from the mode, the second is from the
gapped continuum. At  $-(\Omega_0 + \Delta) > E > -3\Delta$ only
the mode contributes. The self-energy in this range is very
similar to the one-mode result. Above $3\Delta$, the gapped
continuum also contributes to $\Sigma^{''} (E)$, initially as
$\sqrt{E + 3\Delta}/\log^2 (E + 3\Delta)$ for $E \leq -3\Delta$,
and more strongly at larger $|E|$. Combining the contributions
from the mode and from the gapped continuum, we found numerically
that the total $\Sigma^{''}$ is almost flat above $3\Delta$ at a
value $\Sigma^{''} \approx 1.5 \lambda^c_n \Omega_0$. The
near-constant value of $\Sigma^{''}$ is quite close to the normal
state value in the one-mode model, $\Sigma^{''}_n = (\pi/2)
\lambda^b_n \Omega_0$, but we stress that in the collective mode
model, this flat behavior is obtained at $|E| \approx
3\div5\Delta$. In the one-mode model, $\Sigma^{''}(E)$ at these
energies has a strong frequency dependence ranging between $1.25
\Sigma^{''}_n$ at $E = -3\Delta$ to $1.04 \Sigma^{''}_n$ at $E =
-5 \Delta$.

\subsection{Comparison and application of the two models}

Not surprisingly,  one-mode and collective mode models give very
different results for the  normal state. Within the one-mode
model, the normal state spectral function still shows a
peak-dip-hump structure, and the renormalized dispersion displays
an $S$-shape structure near $E = -\Omega_0$. In the collective
mode model, the imaginary part of the self-energy is roughly
linear in $E$ at frequencies comparable to $\Omega_0$, and
$\Sigma^{'} (E)$ displays a crossover from a linear behavior at
small frequencies to a near constant behavior at higher
frequencies. From this perspective, a combination of the
measurements below and above $T_c$ provides the best way to
distinguish between the two models, particularly as we found the
relation between the coupling constants in the normal and
superconducting states. Several ARPES measurements near $(\pi,0)$
indicate~\cite{Kim,Gromko,Sato} that the coupling to the mode
disappears above T$_c$ thus strongly supporting the collective
mode model.

The normal state measurements may be ``contaminated'' by thermal
effects, which  mask the difference between the two models. A  way
to avoid thermal effects is to focus on low $T$ measurements.
However the  two models give very similar results for the
superconducting state. The only qualitative difference is the
gapped continuum which is still present in the collective mode
model, but the continuum affects the self-energy only in a
moderate extent both at vanishing $E$ and at $|E| \sim \Delta$.
The dominant contribution to the self-energy at these energies
comes from the resonance at $\Omega_0$, which is present in both
models. The continuum does affect the self-energy at $|E| \sim
3\div5 \Delta$, but it is difficult to measure the self-energy in
the $(\pi,0)$ region in this energy range  since the bare
dispersion only extends to $\sim 2\Delta$ for the antibonding band
and to $\sim 7\Delta$ for the bonding band and therefore all
evaluations strongly depend on the exact values of the bare
particle dispersion.

On the other hand, the close similarity between the two models in
the superconducting state is good for addressing the fundamental
issue whether the data in the superconducting state are actually
consistent with the collective mode model, and with estimates of
$\lambda^c_{sc}$ or $\lambda^c_{n}$.

For the analysis of the experimental data we used the one-mode
model which is determined by $\lambda^b_{sc}$ and which simulates
the coupling to the magnetic resonance mode.~\cite{neutron} The
bare one-mode model is extended by adding to the self-energy a
Fermi-liquid-like term which is shifted by 3$\Delta$ to higher
binding energy. This term approximates the gapped continuum and is
determined by the coupling constant $\lambda^f_{sc}$. We then
compare the two coupling constants $\lambda^b_{sc}$ and
$\lambda^f_{sc}$ with theoretical estimates for the coupling
constant $\lambda^c_{sc}$ in the collective mode
model.~\cite{Abanov,Abanov3} We obtain reasonable agreement
between experiment and theory which indicates that the dressing of
the charge carriers in the ($\pi$,0) region is related to a
coupling of spin excitations. We also show that the magnitude
$\Sigma''(-\infty)$ evaluated from the incoherent spectral weight
is consistent with the value of $\lambda^b_n$ derived from the
dispersion near $\Delta$. This means that the spectral function
for the coherent and the incoherent states can be described by one
self-energy function indicating a common linear dressing for the
both states.

\section{\label{sec:exper} Experimental setup}
The ARPES experiments were carried out at the BESSY synchrotron
radiation facility using the U125/1-PGM beam line and a SCIENTA
SES100 analyser. Spectra were taken with various photon energies
ranging from 17 to 65 eV. The total energy resolution ranged from
8 meV (FWHM) at photon energies h$\nu$=17-25 eV to 22.5 meV at
h$\nu$=65 eV. The momentum resolution was set to 0.01 {\AA}$^{-1}$
parallel to the $(\pi,0)-(\pi, \pi)$ direction and 0.02
{\AA}$^{-1}$ parallel to the $\Gamma-(\pi,0)$ direction. Here we
focus on spectra taken with photon energies of 38 eV and 50 (or 55
eV) to discriminate between bonding and antibonding bands. The
polarization of the radiation was along the $\Gamma-(\pi,0)$
direction. Measurements have been performed on (1$\times$5)
superstructure-free, optimally doped BiPb2212 single crystal with
a T$_c$= 89 K. Since data for the normal state have already been
published~\cite{Kim} we only show data measured at T = 30 K.

\section{\label{sec:results} Experimental Results}
\begin{figure}
\includegraphics[width=8 cm]{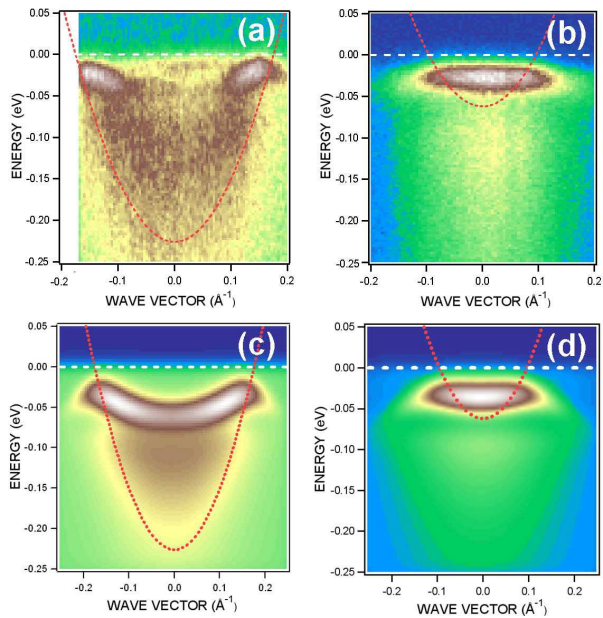}
\caption{ARPES intensity plots as a function of energy and wave
vectors along the $(\pi,\pi)-(\pi,-\pi)$ direction of the
optimally doped Pb-Bi2212 superconductor taken at T = 30 K. Zero
corresponds to the $(\pi,0)$ point. (a) bonding band, (b)
antibonding band. (c) and (d): calculated spectral function using
a model self-energy function for the region around the $(\pi,0)$
point. The red dashed line represents the bare-particle
bandstructure} \label{comp}
\end{figure}

In Fig.~\ref{comp} we show typical ARPES data for wave vectors close to the $(\pi,\pi)-(\pi,-\pi)$ line,
centered around the $(\pi,0)$ point.
As has been shown previously\cite{Borisenko3, Kordyuk5, Kim} and
supported by theoretical calculations\cite{Bansil, Fujimori}, the
data taken with h$\upsilon $=38eV  due to matrix element effects
represent mainly the bonding band with some contributions from the
antibonding band. The data taken at h$\upsilon $=50 eV (see
Fig.~\ref{comp}(b)) have almost pure antibonding character. In
order to obtain the spectral weight of the pure bonding band (see
Fig.~\ref{comp}(a)), a fraction of the 50 eV data has been
subtracted from the 38 eV data. We have also added in
Fig.~\ref{comp} the bare-particle dispersion which was obtained
from a self-consistent evaluation of the data at the nodal
point\cite {Kordyuk3}, an evaluation of the anisotropic plasmon
dispersion\cite {Nucker,Grigoryan} and from LDA bandstructure
calculations\cite {Andersen}. When comparing this bare-particle
dispersion with the very broad distribution of the bonding band at
high energies, one realizes a renormalization of the occupied
bandwidth by a factor of about
1.7 corresponding to $\lambda^f_{sc} \approx 0.7$. This value is
not far from that derived for the bandwidth renormalization above
T$_c$, where no additional renormalization effects at lower
energies have been detected,~\cite {Kim} and from that at the
nodal point~\cite{Kordyuk3, Kordyuk5}. Probably a large fraction
of this bandwidth renormalization stems from a coupling of the
charge carriers to the above mentioned continuum of spin
fluctuations\cite{Abanov}.
\begin{figure}
\includegraphics[width=6 cm]{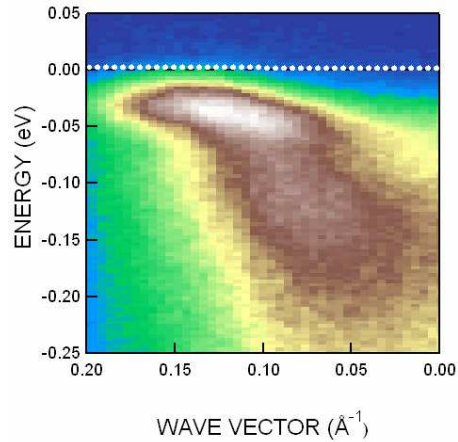}
\caption{ARPES intensity plot for k-values near the
$(1.4\pi,\pi)-(1.4\pi,-\pi)$ line of the optimally doped Pb-Bi2212
superconductor taken at T = 30 K. Zero corresponds to the
$(1.4\pi,0)$ point. The data were taken with a photon energy
h$\upsilon $ = 50 eV in order to maximize the intensity of the
antibonding band.} \label{focus}
\end{figure}
\begin{figure}
\includegraphics[width=6 cm]{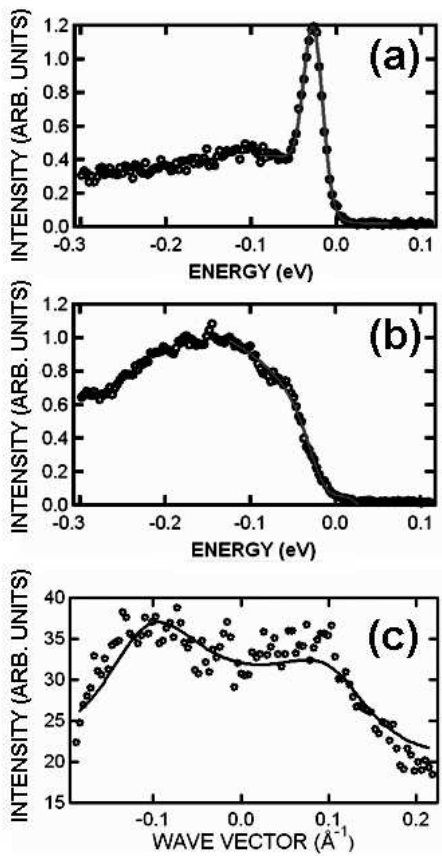}
\caption{(a) Constant-k cut of the data shown in Fig~\ref{focus}
at k$_F$; (b) constant-k cut of the data at about one third of
k$_F$ (starting at the $(\pi,0)$ point); (c) constant-E cut at E =
-100 meV. The solid lines represent fits to the data (see text)}
\label{cuts}
\end{figure}

In Fig.~\ref{focus} we show an ARPES intensity distribution near
k$_F$ of the antibonding band, close to the
$(1.4\pi,\pi)-(1.4\pi,-\pi)$ line, of Pb-Bi2212 measured at 30 K
with a photon energy h$\upsilon$  = 50 eV. At this place in the
second Brillouin zone, the bare particle dispersion of the
antibonding band reaches well below $E_B$ = 70 meV and therefore
contrary to Fig. 4 (b) the branching into two dispersive arms can
be clearly realized. These data together with the data of
Fig.~\ref{comp} when compared with the model calculations shown in
Fig.~\ref{AEk} clearly reveal that the dominant effect of the
renormalization, besides the bandwidth renormalization mentioned
above, is due to a coupling to a bosonic mode leading to a
branching energy of $\sim$ 70 meV.

In order to obtain more quantitative information on the parameters
which determine the self-energy function leading to this
renormalization we have  performed various cuts of the spectral
weight shown in Fig.~\ref{comp} (a) which are presented in
Fig.~\ref{cuts}. A constant-k scan for k=k$_F$ is depicted in
Fig.~\ref{cuts}(a) showing the typical peak-dip-hump structure
presented for data taken at the $(\pi,0)$ point in previous
studies~\cite{Norman, Borisenko3}. From the peak energy one can
derive the superconducting gap energy $\Delta$  = 30(4) meV. In
the previous literature\cite{Eschrig}, from the dip energy at
$\sim$ 70 meV the branching energy E$_B$ was derived. Another
constant-k scan at 1/3 k$_F$ (starting from the $(\pi,0)$ point)
is shown in Fig.~\ref{cuts}(b). At this k-value the intensity of
the coherent peak is strongly reduced and in the framework of the
one-mode model, mainly the threshold of the incoherent states (the
hump) is observed. From a fit to these data, taking into account a
small intensity of the coherent line and a threshold of the
incoherent states, the threshold energy could be determined which
in the one-band model yields the branching energy, $E_B=\Delta
+\Omega _0$ = 70(5)( meV. This together with the gap energy
$\Delta $= 30 meV yields a mode energy of 40 meV.

\begin{figure}
\includegraphics[width=7 cm]{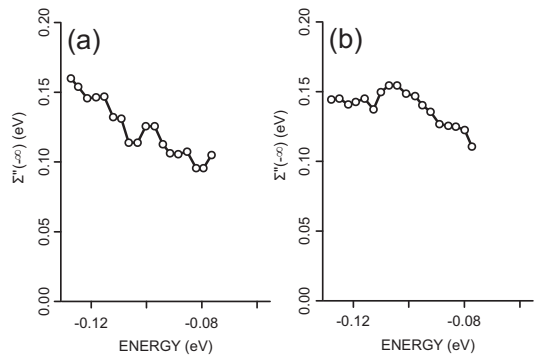}
\caption{Typical values for $\Sigma''(-\infty)$ as derived from a
fit of constant-E cuts (see Fig.~\ref{cuts}) of data shown in
Fig.~\ref{comp}(a) and (b) using Eq. (4) The solid line is a guide
to the eyes. (a) bonding band, (b) antibonding band. }
\label{imsig}
\end{figure}

In Fig.~\ref{cuts}(c) we show a constant-energy cut at $E$ = -100
meV of the data presented in Fig.~\ref{comp}. As discussed in
Sect.~\ref{sec:model}, from the fit of those cuts with a
Lorentzian one can obtain from the width  of the Lorentzian a
value of the imaginary part of the self-energy at the selected
energy. For the energy below the mode energy, this value
$\Sigma''(-\infty)$ is a measure of the coupling to the mode. The
actual situation is more complicated since near the antinodal
point, the bandstructure is far from being linear at this energy
range. We  have fitted the data shown in  Fig.~\ref{cuts}(c) by
Eq. (\ref{26_4}) with the self-energy given by (\ref{25_3}), the
bare dispersion extracted from earlier work,~\cite{Kim}
 $\Delta$ = 30 meV, $\Omega_0$ = 40 meV, and
$\Sigma''(-\infty)$ used as a parameter. Moreover as described
above a Fermi-liquid like term was added to the self-energy to
approximate the influence of the gapped continuum. The imaginary
part of this term is given by $\alpha(E-3\Delta)^2$ for $|E| >
3\Delta$ and zero for $|E| < 3\Delta$ where the magnitude of
$\alpha$ is determined by $\lambda^f_{sc}$. The value $3\Delta $
can be easily understood by looking at Fig.~\ref{boson} since in
the superconducting state the states available for a fermion decay
have minimum energy of $3\Delta$, and hence below $3\Delta$, the
correction to $\Sigma^{''}$ from the continuum is zero. Using this
self-energy function from an extended one-mode model the fit
yielded typical values for the parameter $\Sigma''(-\infty)$ as
shown in Fig.~\ref{imsig}. For $|E| > E_B$ = 70 meV the values
should be constant. The finite slope detected in the analysis may
be related to errors in the bare particle dispersion, to the
assumption of a constant density of states during the definition
of the self-energy function, or to in the assumed
$\lambda^f_{sc}$. For $|E| < E_B$ the results from the fits are
determined by the flat dispersion of the coherent states and
therefore, due to the finite energy resolution, large values are
obtained in this energy range (not shown). From evaluations of
such data taken on several samples we derive a value
$\Sigma''(-\infty)$ = 130(30) meV. The large error for this value
stems from various measurements on samples with slight mismatches
in there orientation which leads to different bare bandstructures
as compared to the assumed one.

Another important information comes from the dispersion of the
coherent spectral weight between the gap energy $- \Delta$ and the
branching energy $-(\Delta + \Omega_0)$. Originally~\cite{Kim,
Gromko}, the data were fitted using Eq. (\ref{a_4}). As pointed
out in Sect.~\ref{sec:model} this is a good approximation for
conventional superconductors, where the mode energy is much higher
than the gap, but not for the high-T$_c$ superconductors. Here the
energy of the mode is comparable to the gap energy and therefore
the renormalization function from which the $\lambda^b_{sc}$
-values are derived, depends on energy and also on $\Delta /\Omega
_0$. In this communication  we have fitted the data using the full
Eqs. (\ref{26_4}) and (\ref{a_3}). For the complex self-energy
function, we have used the same parameters as for the fit of the
constant-energy cuts shown in Fig.~\ref{cuts} (c). We then obtain
a complex renormalization function Z(E) from which we derive
$\lambda$-values not at energies between $\Delta$ and $\Delta
+\Omega$ as in the previous study\cite{Kim} but at zero energy,
i.e., $\lambda^b_{sc}=Z(0)-1$. In the fit we have used the above
given values for $\Delta$ and $\Omega_0$ and we have chosen
$\lambda^b_{sc}$ as a parameter. From the fit, we extracted
$\lambda _{sc}^b=2.0(4)$. This yields a total $\lambda _{sc}^t$=
2.7(5)  from Z(0) in the superconducting state composed of a
bandwidth renormalization part $\lambda _{sc}^f=0.7(3)$ and a
bosonic part $\lambda _{sc}^b=2.0(4)$. The errors are not due to
statistics but result from different measurements on different
samples. Using the relation $\lambda^b_{sc} = 0.74 \lambda^b_n$
(see Sec. ~\ref{sec:model}) we obtain $\lambda^b_n=2.7$. To obtain
$\lambda^f$ in the normal state we used the same $Im \Sigma$ from
fermion-fermion interaction as in the superconducting state, but
set $\Delta $ = 0. This yields $\lambda^f_n \sim 1.3$ and a total
coupling constant for the normal state $\lambda _{n}^t = 4.0(5)$.
We collected the values of the coupling constants and
$\Sigma''(-\infty)$ in Table II. Note that the normal state values
are not derived from measurements in the normal state. Rather they
were derived from data taken in the superconducting state and
setting $\Delta$ to zero in the renormalization function. This
means that $\lambda^b_n$  in Table II is a fictitious normal state
coupling constant because the bosonic mode does not exist in the
normal state. It is only presented for the comparison of the
coupling constant in HTSCs with those of conventional metals and
superconductors.

\begin{table}
\centering \caption{Parameters determining the self-energy
function of Pb-Bi2212 near $((\pi,0)$ below T$_c$.  $\lambda ^f$,
$\lambda ^b$, $\lambda ^t$,   coupling constants  from bandwidth
renormalization, from coupling to a bosonic mode, and total
coupling constant, respectively. $\Sigma''(-\infty)$ is
determining the imaginary part of the self-energy function at high
binding energies. sc: parameters derived in the superconducting
state, n: parameters calculated for the normal state by setting
$\Delta$ to zero. The data are compared with parameter derived for
a Mo(110) surface state \cite{Valla} and for Pb \cite{Reinert}}
\label{tab:2}
\begin{tabular}{lllll}
\hline\noalign{\smallskip}
system &\hspace{.3cm}   $\lambda ^f$&\hspace{.4cm} $\lambda ^b$ &\hspace{.4cm} $\lambda ^t$&\hspace{.4cm}$\Sigma''(-\infty)(meV)$  \\
\noalign{\smallskip}\hline\noalign{\smallskip}
BiPb2212 sc &\hspace{.3cm} 0.7(3) & \hspace{.4cm}2.0(4)& \hspace{.4cm}2.7(5)& \hspace{.4cm}130(30) \\
BiPb2212 n &\hspace{.3cm} 1.3(3) & \hspace{.4cm}2.7(4)& \hspace{.4cm}-& \hspace{.4cm}130(30) \\
Mo(110) &\hspace{.3cm} - & \hspace{.4cm}0.4& \hspace{.4cm}-& \hspace{.4cm}15 \\
Pb n &\hspace{.3cm} - & \hspace{.4cm}1.6& \hspace{.4cm}-& \hspace{.4cm}- \\
\noalign{\smallskip}\hline
\end{tabular}
\end{table}

\section{\label{sec:discussion} Discussion}

The coupling constant to a bosonic mode was derived from the
dispersion of the coherent states between -30 and -70 meV while
$\Sigma''(-\infty)$ was derived from a constant-E cut of the
incoherent states below -70 meV. If the data can be described by
one self-energy function which essentially results from a coupling
to one bosonic mode, the $\lambda^b_n $-value calculated from
$\Sigma''(-\infty)$ within this model should agree with the
$\lambda_n^b$-value extracted from the dispersion of the coherent
states. Using the relation $\lambda _{n}^b =
-2\Sigma''(-\infty)/(\pi \Omega _0)$ (see Sect.~\ref{sec:model})
and the value $\Sigma''(-\infty)$ = 130(30) meV one obtains
$\lambda _{n}^b$ = 2.1(5) which is in reasonable agreement with
$\lambda _n^b \sim 2.7(4)$ derived from the dispersion near
$-\Delta$. This supports the idea that the coherent and the
incoherent spectral weight of the spectral function in the
superconducting state can be described by a {\it single}
self-energy function, which is essentially determined by the
coupling to one bosonic mode at 40 meV. This view is also
supported by the fact that taking this self-energy function and
calculating the spectral function using Eq. (\ref{26_1}) we obtain
a reasonable agreement with the experimental ARPES data, both for
the bonding and the antibonding band (see Fig.~\ref{comp}). The
differences in the intensities of the bonding band near ($\pi$,0)
may be explained by matrix element effects. We emphasize that in
the superconducting state both the real and the imaginary part of
the self-energy function indicate a very strong coupling to a
bosonic mode.

Recently, there has been some evidence from ARPES measurements
\cite{KShen,Roesch} that in the undoped cuprates there is a very
large electron-phonon coupling leading to a strong polaronic
renormalization connected with a negligible spectral weight for
the coherent states and a high spectral weight for a multiphononic
line at higher energies. Furthermore there are theories of the
pairing in high-T$_c$ superconductors which are based on the
formation of polarons and  bipolarons\cite{Alexandrov}. The big
question is whether this strong electron-phonon coupling survives
for the high-T$_c$ superconductors or whether it will be screened
by the charge carriers and at what dopant concentration the
adiabatic approximation is valid, where the Fermi energy is much
larger than the mode energy. The data shown in Fig.~\ref{focus},
which were actually taken down to an energy of -400 meV, show no
indication of a polaronic line at lower energy. Moreover, the data
in Fig.~\ref{focus} and its evaluation in terms of a coupling to a
single bosonic mode gives no room for multi-bosonic polaron
excitations for optimally doped samples.

The analysis given above clearly identifies below T$_c$ a very
strong coupling to a bosonic mode. In Table II we have listed
other coupling constants to bosonic (phonon) modes detected by
ARPES. Compared to the surface state of the Mo(110) surface
coupled to a phonon mode, $\lambda _{n}^b$ in OPBiPb2212  is a
factor of 6 larger. Almost the same factor 8 is obtained for
$\Sigma''(-\infty)$. Compared to the strong coupling
superconductor Pb the coupling constant $\lambda _{n}^b$ for OP
BiPb2212 is a factor of 1.6 larger. This indicates that we really
have a very strong coupling to a bosonic mode. This coupling is
even enhanced at lower dopant concentration \cite{Kim}.

In the following we compare the experimental coupling constants
$\lambda^f_{sc}$ and $\lambda^b_{sc}$ with those derived in the
collective mode model for the gapped continuum and for the single
mode, respectively. In the collective mode model, in the
superconducting state about 20 \% of the total coupling constant
comes from the gapped continuum (see Sec. II B). The corresponding
experimental value $\lambda^f_{sc}/\lambda^t_{sc}$ = 0.26 (see
Table II) is in remarkable agreement with the theoretical value.
This also holds for the absolute values of the coupling constant.
In previous work~\cite{Abanov,Abanov3} the normal state coupling
constant for the collective mode model was estimated to be
$\lambda^c_n = 2\div3$ from fits of theoretical values for
$\omega_{sf}$, $\Omega_0$, and $T_c$. This transforms into
$\lambda^c_{sc} = 1.5\div2.25$ for the superconducting state which
is not far from the experimental value $\lambda^t_{sc}$ = 2.7(5).
Thus the agreement of the relative and absolute experimental
values of the coupling constants with those derived from the
collective mode model is a strong indication that the dressing of
the charge carriers in the ($\pi$,0) region (i.e. the region were
the superconducting order parameter has its maximum) is
predominantly determined by a coupling to spin excitations, in
particular to the magnetic resonance mode.

This interpretation is supported by several other ARPES results.
The strong temperature dependence of the coupling to the
mode~\cite{Kim,Gromko} is difficult to understand in terms of
electron phonon coupling. Our model calculations also show that
the data above T$_c$ cannot be described by a thermally broadened
phonon line. Also the strong dopant dependence~\cite{Kim,Gromko}
is difficult to be explained in terms of electron phonon coupling.
Furthermore there is a large coupling at $(\pi,0)$ and a much
smaller coupling to the mode at the nodal point, indicating a
coupling of states which are separated by a wavevector $(\pi,\pi)$
typical of an antiferromagnetic susceptibility. Moreover the
energy of the bosonic mode detected in ARPES is close to the
energy of the magnetic resonance mode detected in inelastic
neutron scattering. Finally we mention recent ARPES measurements
on the parity of the coupling between bonding and antibonding
band~\cite{BorisenkoP} and the ''magnetic isotope effect'', i.e.,
the strong changes of the dressing of the charge carriers upon
substitution of Cu by Zn.~\cite{Ding, Zabolo}, which both support
the magnetic scenario.

\section{\label{sec:conclusion} Conclusion}

In this contribution we have analyzed the spectral function of
optimally doped BiPb2212 near the antinodal points measured by
ARPES. Compared to previous studies, we have not analyzed just one
constant-k cut or just the dispersion of the coherent state but
the entire spectral function including the coherent and the
incoherent spectral weight. In this context we have used
expressions which not only can be used in the case of normal
superconductors but also for HTSCs where the mode  energy is not
much larger than the superconducting gap. It was possible to
describe the spectral function using a single parameterized
self-energy function. By comparison of the experimental data with
theoretical models, we conclude that the main contribution to the
self-energy is a very strong coupling to the magnetic resonance
mode. At higher energies (and above T$_c$) it was necessary to
take into account a bandwidth renormalization by a factor of two
due to interaction with a gapped (ungapped) continuum of spin
excitations.There is no evidence for multi-bosonic polaron excitations for this dopant concentration.\\

{\bf Acknowledgments}\\

We thank S.V. Borisenko and A.A. Kordyuk  for helpful discussions.
Financial support by the DFG Forschergruppe under Grant No. FOR
538 is acknowledged. One of the authors (J.F.) appreciates the
hospitality during his stay at the Ames Laboratories. A. Chubukov
is supported by NSF-DMR 0240238.


\end{document}